\documentclass[prl,reprint,nobibnotes, aps,showpacs]{revtex4-1}

\usepackage[english]{babel}
\usepackage{grffile}
\usepackage{graphicx,epstopdf,color}
\usepackage{amsmath,amssymb,psfrag}
\usepackage[caption=false]{subfig}

\usepackage[draft]{fixme}
\usepackage[latin1]{inputenc}

\newcommand{\Pin}{P_{\mathrm{in}}}
\newcommand{\Pout}{P_{\mathrm{out}}}
\newcommand{\atil}{\tilde{\alpha}}
\newcommand{\btil}{\tilde{\beta}}
\newcommand{\cATP}{[\text{ATP}]}
\newcommand{\cADP}{[\text{ADP}]}
\newcommand{\cP}{[\text{P$_{\mathrm{i}}$}]}
\newcommand{\Keq}{K_\text{eq}}
\newcommand{\vhd}{v_\text{HD}}
\newcommand{\vld}{v_\text{LD}}
\newcommand{\vmc}{v_\text{MC}}
\newcommand{\fhdmc}{f_{\text{HD}-\text{MC}}}
\newcommand{\jmpbc}{J^{\text{m}}_{\text{PBC}}}
\newcommand{\jcpbc}{J^{\text{c}}_{\text{PBC}}}

\newcommand{\jmld}{J^{\text{m}}_{\text{LD}}}
\newcommand{\jmhd}{J^{\text{m}}_{\text{HD}}}

\newcommand{\phos}{\text{P}_{\text{i}}}

\definecolor{green}{rgb}{0,.3922,0}

\begin{document}

\title{Efficiency at maximum power of interacting molecular machines}
\author{N. Golubeva}
\author{A. Imparato}
\affiliation{Department of Physics and Astronomy, University of Aarhus, Ny Munkegade, Building 1520, DK--8000 Aarhus C, Denmark}
\date{August, 2012}

\begin{abstract}
We investigate the efficiency of systems of molecular motors operating at maximum power. We consider two models of kinesin motors on a microtubule: for both the simplified and the detailed model, we find that the many-body exclusion effect enhances the efficiency at maximum power of the many-motor system, with respect to the single motor case. Remarkably, we find that this effect occurs in a limited region of the system parameters, compatible with the biologically relevant range.
\end{abstract}

\maketitle

Molecular motors are cyclic machines that convert chemical energy into useful work, and perform a variety of cellular functions such as transport, translocation and polymerization \cite{Alberts2007}.

Such machines work in environments at constant temperature, differently from, e.g., heat engines, whose efficiency is bounded by the Carnot's law. Thus, the efficiency of molecular motors is constrained by the thermodynamic limit 1, which can, however, only be achieved in the limit of vanishing power output. Driven by this observation, several researchers have investigated the efficiency of microscopic engines working at maximum power, with the aim of characterizing the optimal balance between these two thermodynamic quantities \cite{Golubeva2012,Seifert2011a,VandenBroeck2012,Esposito2009}. While these works were focused on the efficiency of single motors, little is know on the efficiency of systems of interacting molecular motors. Indeed, molecular devices such as kinesin and myosin move on crowded filaments, where they can encounter other motors, and the resulting interaction can become non-negligible. In fact,  molecular motor traffic is a relevant and widely studied phenomenon, which is typically modelled by using exclusion processes on lattices  \cite{Aghababaie1999,Klumpp2005,Mueller2008,Nishinari2005,Brugues2009,Klumpp2008a,Ciandrini2010,Neri2011,Parmeggiani2003,ASEP}.

In the present letter we investigate the effect of motor-motor interaction on the system efficiency in the maximum power regime, by using two different motor models for kinesin under external mechanical force. The first model is characterized by a single internal conformational state: it is a reduced version of the lattice models that can be found in literature \cite{Kolomeisky2007}, and it is mainly used to discuss in simple terms the issue of EMP in many-body systems.
However, as emphasized in several works \cite{Nishinari2005,Brugues2009,Ciandrini2010,Klumpp2008a}, it is crucial to include internal states if one wants to characterize properly the mechano-chemical cycles of molecular devices. Thus, we exploit the 6-state  model for single kinesin introduced in \cite{Liepelt2007}, and extend it in order to investigate the exclusion interaction effect on several motors moving on a lattice at the same time.

For both the models we find the striking effect that the EMP is enhanced, as compared to the single motor case, by the mutual exclusion interaction. The increase in the EMP is mainly driven by the dynamical phase transition that the system undergoes as the mechanical force is increased. Most remarkably, the increase of the EMP occurs in a narrow range of parameters, corresponding to the biologically significant one.  

\begin{figure}
 \psfrag{emp}[ct][ct][1.]{$\eta^*$}
 \psfrag{Dmu}[ct][ct][1.]{$\Delta\mu$}
 \psfrag{legend1}[ct][ct][1.]{$\theta=0.3$}
 \psfrag{legend2}[ct][ct][1.]{$\theta=0.65$}
 \psfrag{MC}[ct][ct][1.]{{\footnotesize MC}}
 \psfrag{HD}[ct][ct][1.]{{\footnotesize HD}}
 \psfrag{Dmu1}[ct][ct][1.]{{\scriptsize $\Delta\mu=10T$}}
 \psfrag{Dmu2}[ct][ct][1.]{{\scriptsize $\Delta\mu=18T$}}
 \psfrag{Dmu3}[ct][ct][1.]{{\scriptsize $\Delta\mu=23T$}}
 \psfrag{pout}[ct][ct][1.]{$\Pout$ (pN nm/s)}
 \psfrag{f}[ct][ct][1.]{$f$ (pN)}
 \psfrag{a}[cB][cB][1.]{{\scriptsize $\textbf{(a)}$}}
 \psfrag{b}[cB][cB][1.]{{\scriptsize $\textbf{(b)}$}}
 \psfrag{c}[cB][cB][1.]{{\scriptsize $\textbf{(c)}$}}
 \psfrag{d}[cB][cB][1.]{{\scriptsize $\textbf{(d)}$}}
\centering
\includegraphics[width=\columnwidth]{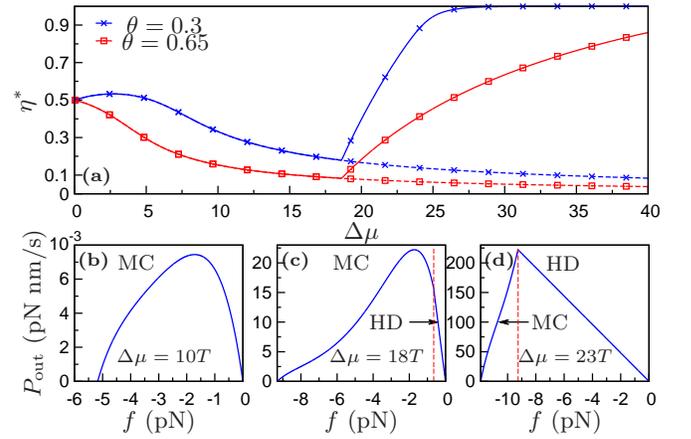}
\caption{Model I: a) EMP  for a single-state model of kinesin interacting through self-exclusion (solid lines) for two different values of the load distribution factor $\theta$. For comparison, the EMP for non-interacting motors with the same parameter values is shown with dashed lines. Parameter values: $a=8$ nm, $\omega_0=1.3\cdot 10^{-7}$ s$^{-1}$, $\alpha=5$ s$^{-1}$, $\beta=3$ s$^{-1}$. The values for $\theta$ are taken from \cite{Liepelt2007} and correspond to two independent experiments on the kinesin motor. b)--d) $\Pout$ as a function of $f$ for three different values of the chemical free energy $\Delta\mu$ and for $\theta=0.3$. The vertical dashed line in c)--d) represents the phase boundary between the HD and the MC phase.
}
   \label{fig:modelI}
 \end{figure}

\paragraph{Model I: Model with no internal states.} The simplest description of a system of interacting molecular machines moving on a filamentous track is obtained by neglecting the internal conformational changes of the motor, and modelling the stepping of an individual motor by a single Poissonian step. The stepping is subject to an exclusion rule similar to the widely studied asymmetric simple exclusion process (ASEP) \cite{ASEP}. The motor is thus described as a particle moving on a lattice with a lattice constant $a$ corresponding to the step size of the motor. The forward and backward  jumping rates are denoted by $p$ and $q$, respectively. In the absence of interactions the motor will thus exhibit a steady-state velocity $v_0=a(p-q)$. However, when the exclusion rule is introduced, the step is rejected, if the motor attempts to step into a site already occupied by another motor, and the steady-state velocity $v$ of a single motor is smaller than $v_0$. The particles bind to the filament at the left end with rate $\alpha$ and leave the track at the right end with rate $\beta$.
The motor performs work against the external load force $f<0$ while hydrolyzing one ATP molecule into one ADP and one $\phos$ molecule during each step. For constant concentrations of ADP and $\phos$, we can write the transition rates according to chemical kinetics as  
\begin{equation}
\label{eq:rates}
 p=\omega_0 e^{(\Delta\mu+f a \theta)/T}, \qquad  q=\omega_0 e^{-f a (1-\theta)/T},
\end{equation} 
where $\omega_0$ is a microscopic rate, $\theta$ is the load sharing factor, expressing the coupling between the force and the system kinetics, and $T$ is the temperature, which is taken to be $T=4.1$ $\text{pN}\cdot\text{nm}$. Here and in the following we take $k_B=1$. The quantity $\Delta\mu$ is the chemical free energy of ATP hydrolysis, which can be expressed in terms of the reactant concentrations: $\Delta\mu=T\ln\{ \Keq \cATP/(\cADP \cP) \}$, where $\Keq$ is the equilibrium constant of the reaction.  
Thus, in this simplified model motor, the mechanical and the chemical cycles are tightly coupled, and an ATP molecule is hydrolysed only when a mechanical step takes place.
The velocity of a single motor is equal to $v=J/\rho$, where $J$ is the steady-state probability current of the particles on the lattice, and $\rho$ is the (average) bulk density of the motors. The quantities $J$ and $\rho$ can be obtained exactly, and the phase diagram of the probability current in the thermodynamic limit is that of a standard ASEP \cite{ASEP}, see also App. A1 in \cite{appendix}. When projected onto the effective parameters $\atil=\alpha/(p-q)$ and $\btil=\beta/(p-q)$, it consists of three regions termed low-density (LD), high-density (HD) and maximal current (MC) phase, respectively. The current is given by $J=a\rho(1-a\rho)$, where $\rho$ depends on the phase, and its value can also be obtained within several mean-field approaches \cite{ASEP,appendix}. In the LD phase, characterized by $\atil<\min(\btil,1/2)$, the density is $\rho=\atil/a$, while for $\btil<\min(\atil,1/2)$ the system is in the HD phase and $\rho=(1-\btil)/a$. Finally, in the MC phase, characterized by $\atil>1/2$ and $\btil>1/2$, the bulk behaviour becomes independent of boundary conditions and $\rho=1/2a$.  Hence, the velocity of the interacting motors becomes
\begin{equation}
\label{eq:v_asep}
v=
\begin{cases}
v_0-a\alpha & \text{for $\atil<\frac{1}{2}$, $\btil>\atil$ (LD)} \\
a\beta & \text{for $\btil<\frac{1}{2}$, $\btil<\atil$ (HD)} \\
v_0/2 & \text{for $\atil>\frac{1}{2}$, $\btil>\frac{1}{2}$ (MC)}.
\end{cases}
\end{equation}
It is worth noting that while the total particle current $J$ is maximized in the MC phase, the single particle velocity $v$ attains its maximum in the LD phase as expected, since  the hindrance due to the other particles is minimal in this phase. Moreover, the velocities fulfill the relation $v_0>\vld>\vmc>\vhd$. In the HD phase the velocity is constant due to queuing and only depends on the detachment rate $\beta$.  
Finally, we note that the typical microtubule lengths $L\sim 10 \mu $m \cite{Howard2001} are long compared to the motor step size $a=8$~nm, and the thermodynamic limit is thus applicable. 
\paragraph{Efficiency at maximum power.} We proceed by evaluating the EMP for model~I.
We fix the value of $\Delta \mu$ (which corresponds to fixing the ATP concentration as discussed above), and look for the optimal load force $f^*$ for which the output power $\Pout=-fv$ is maximum, i.e. $f^*$ is solution of $\partial \Pout/\partial f =-(v+f \partial v/\partial f)=0$, where $v$ is determined from eq. \eqref{eq:v_asep}. It is worth noting that changing $f$ changes the values of the forward and backward rates $p$ and $q$, as given by eq.~(\ref{eq:rates}), and thus the values of $v_0$ as well as $\atil$ and $\btil$. This in turn may induce a phase transition, according to the rules in eq.~(\ref{eq:v_asep}), as will be discussed below. Since the motor is operating in the tightly coupled regime, the efficiency is simply given by $\eta=-\Pout/\Pin=-fa/\Delta\mu=f/f_s$, where $\Pin=\Delta\mu v/a$ is the input power, and $f_s=-\Delta\mu/a$ is the stall force. Therefore, the EMP for a given value of $\Delta\mu$ (and thus of $f_s$) reads $\eta^*=f^*/f_s$. We then repeat this procedure for increasing values of $\Delta \mu$.
We choose the values of the four parameters $\omega_0$, $\theta$, $\alpha$ and $\beta$ as reported in fig.~\ref{fig:modelI} caption. Such values are suitable for the kinesin motor, see App. A2 \cite{appendix} for a detailed discussion.

The resulting EMP is shown in fig. \ref{fig:modelI}a, where two different regimes for the behaviour of such a quantity are observed. When the free energy of hydrolysis is small, the motor operates at maximum power in the MC phase, see fig. \ref{fig:modelI}b. Since the velocity in this regime is $v=\vmc=v_0/2$, the resulting EMP is identical to that obtained for the single motor in the absence of interaction. In particular, we recover the well-known linear response value for the EMP, $\eta^*=1/2$, as $\Delta\mu/T\to 0$ \cite{Golubeva2012,Seifert2011a,VandenBroeck2012}. By increasing $\Delta\mu$ further, the system is in the HD phase at small $f$, but then the maximal $\Pout$ is achieved in the MC phase, see fig. \ref{fig:modelI}c.
However, there is a critical value $\Delta\mu_c$ for the free energy, that is achieved when the velocity at maximum power in the MC phase is equal to the velocity in the HD phase, the latter quantity being independent of $\Delta\mu$ and $f$, see eq.~(\ref{eq:v_asep}). For the present choice of parameters one finds  $\Delta\mu_c\simeq 18.5\, T$, see App. A2 in \cite{appendix}. When $\Delta\mu>\Delta\mu_c$, the maximum power is no longer obtained when the system is in the MC phase, but rather at the boundary between the two phases, and the load force maximizing $\Pout$ thus reads $f^*=\fhdmc$, see figure \ref{fig:modelI}d. Since the velocity of non-interacting motors, $v_0(f)$, decreases with the force for \emph{all} values of $f<0$, one finds that the optimal force $f_0^*$ is smaller than $f^*$ when $\Delta\mu>\Delta\mu_c$. Thus, for $\Delta\mu>\Delta\mu_c$, we observe an increase in the EMP, $\eta^*=f^*/f_s$, as compared to the non-interacting system, due to a change in the characteristic force-velocity relation $v(f)$. Furthermore, the transition force $\fhdmc$ is such that $\fhdmc \to f_s$ as $\Delta\mu/T$ goes to infinity. Hence, $\eta^*=f^*/f_s\to 1$ for $\Delta\mu/T\to\infty$ entirely due to the self-interaction of the motors on the lattice. We note, however, that the limit $\eta^* \to 1$ for $\Delta\mu/T \to \infty$ is unphysical, since in real systems the velocity does not grow unbounded with increasing $\Delta\mu$ because of the dissipation. One would rather expect the EMP to decrease for very large $\Delta\mu$, as we will see for the more realistic model II. However, large values for the EMP, $\eta^* \gtrsim 0.6$, are achieved for $\Delta\mu$ moderately larger than $\Delta\mu_c$, before the saturation begins to take place.

We find that the enhancement of the efficiency at maximum power, as compared to the EMP for non-interacting motors, is primarily determined by the load factor $\theta$, as shown in fig.~\ref{fig:modelI}a, and as discussed in App. A2 in \cite{appendix}. It is interesting to note, that for the value of the chemical free energy under physiological conditions, $\Delta\mu/T \approx 20-25$ \cite{Howard2001}, the model predicts an increase in the EMP up to  a factor of four. 

\begin{figure}
   \centering
   \includegraphics[width=0.8\columnwidth]{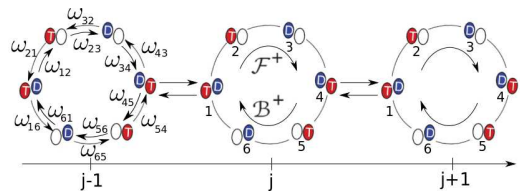}
   \caption{Model II: The 6-state network model from \cite{Liepelt2007} for a single kinesin motor is adapted to describe kinesin traffic by introducing self-exclusion. States marked by $D$ or $T$ represent a head domain containing bound ADP or ATP, respectively, while an empty circle denotes an empty head domain. The motor uses ATP hydrolysis to perform a forward mechanical step from site $j$ to site $j+1$ through the $\mathcal{F}^+=|12341 \rangle$ dicycle, or a backward step to site $j-1$ through $\mathcal{B}^+=|45614 \rangle $, if the respective neighbouring site is unoccupied. Furthermore, the network contains the futile hydrolysis dicycle $| 1234561 \rangle$.} 
   \label{fig:LC}
 \end{figure}

\paragraph{Model II: Model with internal states.}
We now consider the 6-state network model for single kinesin motors introduced in ref. \cite{Liepelt2007}, and depicted in fig.~\ref{fig:LC}, and extend the model to describe traffic of kinesin motors interacting through steric interactions as above. 
The exclusion rule affects the transitions between chemical states 1 and 4 since these transitions represent mechanical steps between neighbouring sites. As in model I, the motors can attach to the microtubule at the left boundary with rate $\alpha$ and detach at the right boundary with rate $\beta$.
The dynamics of the system can be obtained analytically within a mean-field approximation by employing the maximal current principle (MCP) \cite{MCH}. 
The MCP states that the mechanical probability current through the open system in the thermodynamic limit, $J^{\text{m}}$, is given by \cite{MCH}
\begin{equation}
  \label{eq:mch}
  J^{\text{m}}=
\begin{cases}
\max \limits_{\rho \in [\rho_r,\rho_l]} \jmpbc(\rho) & \text{for $\rho_l>\rho_r$} \\
\min \limits_{\rho \in [\rho_l,\rho_r]} \jmpbc(\rho) & \text{for $\rho_l<\rho_r$}, 
\end{cases}
\end{equation}
where $\rho_l$ and $\rho_r$ are the densities of the left and right reservoirs, respectively. The mechanical current in the bulk, $\jmpbc(\rho)$, is obtained by solving the mean-field master equations for the corresponding homogeneous system with periodic boundary conditions (PBC), and homogeneous particle density $\rho$, see eq.~(14) in App. B2 \cite{appendix}. We find that $\jmpbc$ has a single maximum, and according to the MCP the topology of the phase diagram is therefore identical to that of a standard ASEP with the three phases as in model I, as can be seen in fig. 2 in App. B2 \cite{appendix}. The reservoir densities $\rho_{l,r}$ as a function of the transition rates $\omega_{ij}$ and the injection and removal rates $\alpha$ and $\beta$ are obtained by equating the bulk current $\jmpbc$ with the particle currents at the boundaries, i.e.
$\jmpbc(\rho_l)=\alpha(1-a \rho_l)$, and  $\jmpbc(\rho_r)=\beta a \rho_r$. Here we have assumed that the motor can detach from the track from any internal state. 

The maximal current principle yields the bulk densities $\rho=\rho_l$ in the LD phase, $\rho=\rho_r$ in the HD phase, and $\rho=\rho^*$ in the MC region, where $\rho^*=\max_\rho\jmpbc(\rho)$ denotes the density that maximizes the current under PBC. The mechanical and chemical probability currents are given by $J^{\text{m}}=\jmpbc(\rho)$ and $J^{\text{c}}=\jcpbc(\rho)$, respectively, where $\rho$ is the appropriate bulk density for the phase, and $\jcpbc$ is the chemical probability current through the homogeneous system. The phase boundary between the LD and MC phases given by $\alpha_c$ is obtained by solving $\rho_l(\alpha_c)=\rho^*$, while the critical value $\beta_c$ defining the boundary between the HD and MC phases is determined by the relation $\rho_r(\beta_c)=\rho^*$. Finally, the LD-HD phase boundary is defined by the equation $\jmld(\alpha)=\jmhd(\beta)$ that yields a linear relation, $\beta=\alpha \beta_c/\alpha_c$. For fixed $\alpha$ and $\beta$, and for given values of $\Delta\mu$ and $f$, we can thus calculate the phase boundaries analytically and determine the phase that the motor operates in under these conditions, and hence $J^m$ and $J^c$. One thus obtains the motor velocity $v=J^m/\rho$ and the hydrolysis rate $r=J^c/a\rho$.

\begin{figure}
 \psfrag{etas}[ct][ct][1.]{$\eta^*$}
 \psfrag{Dmu}[ct][ct][1.]{$\Delta\mu$}
 \psfrag{legendlegendlegendlegend1}[ct][ct][1.]{{\footnotesize $\chi_1=0.05,\chi_2=0.25$}}
 \psfrag{legendlegendlegendlegend2}[ct][ct][1.]{{\footnotesize $\chi_1=0.3, \chi_2=0.4$}}
 \psfrag{MC}[ct][ct][1.]{{\footnotesize MC}}
 \psfrag{HD}[ct][ct][1.]{{\footnotesize HD}}
 \psfrag{LD}[ct][ct][1.]{{\footnotesize LD}}
 \psfrag{Pout}[ct][ct][1.]{$\Pout$ (pN nm/s)}
 \psfrag{f}[ct][ct][1.]{f (pN)}
 \psfrag{eta}[ct][ct][1.]{$\eta$}
 \psfrag{a}[ct][ct][1.]{{\scriptsize $\textbf{(a)}$}}
 \psfrag{b}[ct][ct][1.]{{\scriptsize $\textbf{(b)}$}}
 \psfrag{c}[ct][ct][1.]{{\scriptsize $\textbf{(c)}$}}
 \psfrag{la}[ct][ct][1.]{{\footnotesize \textcolor{green}{$\eta^*$}}}
 \psfrag{lb}[ct][ct][1.]{{\footnotesize \textcolor{green}{$\eta_0^*$}}}
   \centering
   \includegraphics[width=\columnwidth]{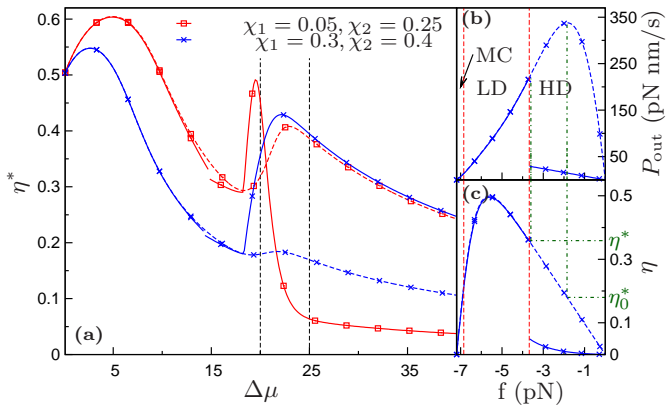}
   \caption{Model II: a) The EMP for interacting (solid lines) and non-interacting (dashed lines) kinesin motors as a function of the hydrolysis  free energy $\Delta\mu$ for two different parameter sets. Squares: parameters obtained in \cite{Liepelt2007} for the Visscher et al. experiment \cite{Visscher1999}. Crosses: as before but with chemical load parameters $\chi_{1,2}$ as in \cite{Liepelt2010} (see text). For all curves we take the biological concentrations $\cADP=0.07$ mM, $\cP=1$ mM \cite{Hackney2005} and vary the ATP concentration. Where relevant, we take $\alpha=5$ s$^{-1}$ and $\beta=3$ s$^{-1}$. The biologically relevant regime $\Delta\mu/T \approx 20-25$ is marked with vertical lines. The discontinuity in the EMP for interacting motors at $\Delta \mu\simeq 14.5 T$ arises due to a discontinuity in the maximizing force $f^*$ as a result of the MC-LD phase transition. b) $\Pout$ as a function of $f$ for the second parameter set and for $\Delta\mu=20T$. As $f$ increases, the motor goes from operating in the HD phase to the LD phase, while the MC phase is entered close to stall force. The maximum power is achieved at the HD-LD boundary. c) The efficiency $\eta(f)$ corresponding to the parameters as in b). The shift of $f^*$ to higher values results in a boost of the EMP, $\eta^*$, with respect to the non-interacting case, $\eta^*_0$.} 
   \label{fig:etaII}
 \end{figure}

\paragraph{Efficiency at maximum power.}
Since the motor is loosely coupled due to the presence of backward hydrolysis-driven stepping and futile hydrolysis, the efficiency is now given by $\eta=-fv/\Delta\mu r$, and generally $f_s<\Delta\mu/a$. The efficiency at maximum power predicted for model II is shown in fig.~\ref{fig:etaII}a for two different sets of parameters. The difference between these two sets consists only in different values for the chemical load factors $\chi_{1,2}$, which express the coupling between the external mechanical force $f$, and the chemical transition rates, see ref.~\cite{Liepelt2007} and App. B1 in \cite{appendix}. 
For the parameters obtained in \cite{Liepelt2007} (squares in fig.~\ref{fig:etaII}a) by fitting the Visscher et al.'s experimental data \cite{Visscher1999}, we observe that the EMP is enhanced significantly, as compared to the EMP for non-interacting motors under the same conditions, when the input free energy is in the range $\Delta\mu/T \approx 18-21$  \cite{note_param}. The mechanism for the EMP enhancement is well exemplified by fig. \ref{fig:etaII}b where we compare $\Pout$ for interacting and non-interacting motors. At low force the interacting system is in the HD phase, while at intermediate force we observe a phase transition to the LD phase, and our model molecular machines operate at maximum power at the transition between the two phases. Thus the force $f^*$ maximizing $\Pout$ is shifted to higher values with respect to the non-interacting system values $f_0^*$ because of the change in the force-velocity relation $v(f)$. Since the efficiency, $\eta(f)$, in the LD phase is approximately equal to the efficiency for the non-interacting motors, $\eta_0(f)$, the higher value for $f^*$ implies an increase in the efficiency at maximum power as long as $\Delta\mu \lesssim 21T$, as shown in fig.~\ref{fig:etaII}c. When $\Delta\mu \approx 21T$, the optimal force $f^*$ becomes so large that $\eta^*=\eta(f^*)=\eta_0(f_0^*)$. For even higher values of $\Delta\mu$, the value of $f^*$ approaches the stall force, while the input energy is dissipated through the futile cycles, and hence $\eta^* \to 0$ as $\Delta\mu \to \infty$.   

The range of values for the free energy $\Delta\mu$ for which the EMP exhibits an enhancement is seen to partly overlap with the biologically applicable regime $\Delta\mu/T \simeq 20-25$ \cite{Howard2001}. We find that the width of the ``peak'' that determines the region for the increase in the EMP, is very sensitive to the values of the chemical load parameters $\chi_{1,2}$, and not to variations of the other model parameters as discussed in App. B3 \cite{appendix}. As a second set of parameters we take the values $\chi_1=0.3$ and $\chi_2=0.4$, as obtained in \cite{Liepelt2010} by fitting the data from the Carter et al.'s experiment \cite{Carter2005} to a 7-state model that is an extension of the 6-state model considered in the present work. We plot in fig. \ref{fig:etaII}a the resulting EMP for these values, while keeping all the other parameters fixed (crosses). In this case, the EMP exhibits approximately a two-fold increase in the presence of interactions for a wide range of free energies, $\Delta\mu/T \approx 18-40$, with a wider overlap with the biologically relevant regime $\Delta \mu/T\simeq 20-25$.
We note that the values for the  parameters $\chi_i$  exhibit a strong dependence on the underlying model used to fit the experimental data \cite{Liepelt2007,Liepelt2010}. Furthermore, the parameters $\chi_i$ depend strongly on the experimental conditions: in ref. \cite{Liepelt2007}, by fitting the same model to two different experimental data sets, \cite{Visscher1999} and \cite{Carter2005}, $\chi_1$ takes the values $0.05$ and $0.15$, respectively. We note that other factors may also contribute to shifting the peak of EMP to higher values of $\Delta\mu$ as discussed in App. B3 \cite{appendix}. 


\paragraph{Conclusions.} 
We have studied the EMP for two different models of kinesin traffic, and  showed that the motor-motor interaction plays a fundamental thermodynamic role by enhancing the system efficiency at maximum power, as compared to the single motor case. Our findings are amenable of experimental verification, given that a few experimental groups have started measuring the dynamical properties of many-motor systems \cite{Leduc2012,Seitz2006}, which can be easily compared to the single-motor ones. 
We note that when multiple motors are involved in transporting a single cargo, the force-velocity characteristic is modified, as compared to the single-motor-single-cargo system \cite{Guerin2010,Klumpp2005,Mueller2008}. However, in a recent \emph{in vivo} study \cite{Leidel2012}, Leidel et al. conclude that the motors share the load force equally. Based on this, one can argue that groups of motors bound to several fluid cargos may exhibit a dynamical phase transition as the one discussed above, leading to a similar enhancement of the EMP.
Finally, we expect our findings to hold beyond the case of molecular motors on a lattice: in a generic system of mechanochemical motors, interacting with a general potential, one might presumably observe an enhancement in the EMP, provided that the system exhibits a dynamical phase transition between two regimes with different characteristic response to the external mechanical driving. This is worthy of future investigation. 
\acknowledgements
We gratefully acknowledge financial support from Lundbeck Fonden, and from the ESF network "Exploring the physics of small devices".

\bibliography{interaction4_cm.bbl}

\end{document}